
\input harvmac
\noblackbox

\def\ie{{\it i.e.}}
\def\Dslash{\raise.15ex\hbox{/}\kern-.77em D}

\def\CE{{\cal E}}
\def\ex#1{{\rm e}^{^{#1}}}
\Title{\vbox{\hbox{PUPT-1386}
\hbox{NSF-ITP-93-66}
\hbox{SNUTP 93-27}
\hbox{\tt hep-lat@xxx/9305026}
}}
{\vbox{\centerline{3 Into 2 Doesn't Go:}
\vskip2pt\centerline{\titlerms (almost) chiral gauge theory}
\vskip2pt\centerline{\titlerms on the lattice$^*$
}}}
\footnote{}{{\parindent=-10pt\par $\star$
\vtop{
\hbox{Email: {\tt distler@puhep1.princeton.edu,\quad
sjrey@phya.snu.ac.kr} .}
\hbox{Research supported by NSF grants PHY90-21984, PHY89-04035,
DOE grant DE-FG02-91ER40671, }
\hbox{KOSEF-SRC and Ministry of Education.}
} }}
\vskip-0.5cm
\centerline{Jacques Distler$^{a,b}$ \& Soo-Jong Rey$^{a,c}$}
\bigskip
\centerline{\baselineskip=7pt
\vbox{\sl
\hbox{Joseph Henry Laboratory$^a$}
\hbox{Princeton University}
\hbox{Princeton, NJ 08544 USA}
}
\quad
\vbox{\sl
\hbox{Institute for Theoretical Physics$^b$}
\hbox{University of California}
\hbox{Santa Barbara, CA 93016 USA}
}
\quad
\vbox{\sl
\hbox{Center for Theoretical Physics$^c$}
\hbox{Seoul National University}
\hbox{Seoul 151-742 KOREA}
}
}
\vskip .3in
Kaplan recently proposed a novel lattice chiral gauge theory in which
the bare theory is defined on $(2n+1)$-dimensions, but the continuum
theory emerges in $2n$-dimensions. We explore whether the resulting
theory reproduces all the features of continuum chiral gauge theory
in the case of two-dimensional axial Schwinger model. We find that
one can arrange for the two-dimensional perturbation expansion to be
reproduced successfully. However, the theory {\it fails} to reproduce
the 2-dimensional fermion number nonconservation.
\vskip-1cm
\Date{5/93} 
\vfill\eject

\lref\kaplan{D. Kaplan, Phys. Lett. \bf 288B \rm (1992) 342,
{\tt hep-lat/9206013}.}
\lref\jensen{K. Jensen and M. Schmaltz, UCSD-PTH 92-29 (1992).}
\lref\golterman {M. Golterman, K. Jensen and
                 D. Kaplan, UCSD-PTH 92-28 preprint (1992).}
\lref\callanharvey {C.G. Callan and J.A. Harvey, Nucl. Phys.
\bf B250 \rm (1985) 427.}
\lref\nielsenninomiya {H.B. Nielsen and M. Ninomiya, Nucl. Phys.
\bf B185 \rm (1981) 20;
Karsten and J. Smit, Nucl. Phys. \bf 183 \rm (1981) 103.}
\lref\banks{T. Banks,
Phys. Lett. \bf 272B \rm (1991) 75.}
\lref\dabholkar{T. Banks and  A. Dabholkar, Phys. Rev. {\bf D46} (1992)
4016, {\tt hep-lat/9204017}.}
\lref\wilsonyukawa{J. Smit, Nucl. Phys. \bf B175 \rm
(1980) 307; Acta Phys. Pol. \bf B17 \rm (1986) 531; P.D.V. Swift, Phys.
Lett. \bf 145 B \rm (1984) 256.}
\lref\eichtenpreskill{E. Eichten and J.P. Preskill,
Nucl. Phys. \bf B768 \rm (1986) 179.}
\lref\staggered{J. Kogut and L. Susskind, Phys. Rev. \bf D11 \rm (1975) 395;
L. Susskind, Phys. Rev. \bf D16 \rm (1977) 3031; see also J. Smit, talk at
Lattice `87, Seillac, France (1987).}

\newsec{Introduction}
Under several reasonable assumptions to lattice gauge theory,
such as translation-~, gauge-~, chiral-invariant, local and
quadratic Hamiltonian,
and an implicit assumption that the
Higgs dynamics is unimportant, Nielsen and Ninomiya, and Karsten and
Smit \nielsenninomiya\ proved
that any attempt to put a chiral fermion on the lattice
is afflicted
with unremovable doubler fermions of opposite chirality.

Kaplan recently proposed a clever method  \kaplan  for
defining chiral fermions
on the lattice, which might evade the above Nielsen-Ninomiya No-Go theorem.
His starting point is a $d=(2n+1)$-dimensional Wilson fermion
with a Yukawa coupling to a $2n$-dimensional \sl static \rm
topological defect. At low-energy below a threshold determined by
Wilson mass and mass gap of the topological defect \jensen
the effective theory consists of massless chiral fermions
trapped to the topological defect, while the `would-be' doubling
modes are naturally made heavy. The chiral zero mode currents are
supplied by nontrivial $d$-dimensional lattice
Chern-Simons currents \golterman
-a lattice version of the effect discovered by Callan and Harvey
\callanharvey.

Kaplan's proposal is not complete for defining a full-fledged
chiral gauge
theory on the lattice, however. One has to ensure that gauge
field dynamics \sl also \rm reduces from the microscopic
$d=(2n+1)$-dimensional theory to a low-energy $2n$-dimensional
effective theory. Only if this `dynamical dimensional
reduction' is realized in a
way compatible with the Kaplan's lattice chiral fermions,
a truly well-defined prescription of putting a chiral gauge theory
on the lattice is accomplished. It is this issue we would like to
address in this paper, in the simplest context of 3 into 2 dimensions
and an abelian gauge theory.

This paper is organized as follows. In the next section, we
briefly explain Kaplan's proposal of lattice chiral fermion
and anisotropic gauge theory in $2n+1$ dimensions.
In section 3, we recapitulate essential features of the
two-dimensional axial Schwinger model.
In section 4, the three-dimensional lattice gauge theory is shown
to factorize, at the perturbative level,
into a product of two two-dimensional chiral gauge theories at
low energies: one on the wall and the other on the anti-wall.
Infinite volume limit is then expected to isolate a chiral
gauge theory induced on the wall or the anti-wall.
Despite this success, in section 5,
we argue that anomalous global symmetry such as fermion
number violation is not reproduced and, in fact, completely suppressed
in the infinite mass gap limit.

\newsec{Kaplan's Proposal for Lattice Chiral Fermion}
Kaplan defines a lattice Dirac fermion starting from a one higher
dimensions: $d=(2n+1)$-dimensional
Euclidean lattice $x_\mu \equiv ({\bf x}, x_{2n+1} \! \equiv \! s)$.
In any concrete realization of his proposal, we must actually deal with
a finite lattice with (periodic or antiperiodic) boundary conditions on
the fields. The periodicity in the ${\bf x}$-directions will not be of
much concern to us, but the periodicity in the $x_{2n+1}$-direction does
present several new features, which will be important to us. For
concreteness, we will take $s\in[-L,L]$, with periodic boundary
conditions on the fields.
The lattice action includes both a Wilson mass term
and
a $(2n+1)$-th coordinate-dependent Dirac
mass term:
\eqn\faction{
S_{\rm fermion} =
\sum_{x \in {\bf R}_d } \bar \Psi_x [{\bf K} + {\bf
M} (s)] \Psi_x
}
in which
\eqn\Kdef{
({\bf K} \Psi)_x \equiv  \,\, {1 \over  a} \sum_{\hat \mu = 1}^{2n+1}
\half\gamma^\mu (\Psi_{x + \hat \mu} - \Psi_{x - \hat \mu})
+ ra (\Psi_{x + \hat \mu} - 2 \Psi_x + \Psi_{x - \hat \mu})\quad.
}
When the $x_{2n+1}$-direction is uncompactified, we can simply take the
position-dependent mass term to be of the form of a ``kink" centered at
$s=0$
\eqn\Mdef{
{\bf M}(s)\equiv {\bf M}_0(s)=
{sinh (a M_d) \over a} (\theta (s) - \theta
(-s)).}
However, in a finite volume, the mass term must  be a {\it periodic}
function of $s$, ${\bf M}(s)={\bf M}(s+2L)$, so we will take it to
consist of a kink at $s=0$, and an ``anti-kink" at $s=L$
\eqn\eMdefb{
{\bf M}(s)=\half\bigl({\bf M}_0(s)-{\bf M}_0(s-L)\bigr)\quad.}
It is, perhaps, most elegant to imagine that the position-dependent mass
term ${\bf M} (s)$ arises dynamically from
a Yukawa coupling to a scalar field whose vacuum expectation value is
that of a (pair of) $2n$-dimensional topological defect(s). However,
adding a scalar field to the theory only complicates the dynamics
without adding any new desirable features. It is simpler, and this is
the path we will follow, to just put
in the position-dependent mass {\it by hand}.

As in the continuum theory \callanharvey, the defects trap fermion zero modes.
The expressions for the zero modes are easiest to write down in the
absence of the Wilson term ($r=0$). On a periodic lattice,
\eqn\ezeromodes{
\Psi_{LH}({\bf x},s)=C\e{-M_d f(s)}{\psi({\bf x})\choose0},\qquad
\Psi_{RH}({\bf x},s)=C\e{-M_d f(s-L)}{0\choose\chi({\bf x})}
}
where $f(s)$ is  the periodic sawtooth function,
$f(s)=(|s|-|s+L|+|s+2L|-L)\equiv f(s+2L)$ and
$$C=\sqrt{M_d}\left(1-\ex{-2M_dL}\right)^{-1/2}\quad.$$
On the infinite lattice, only
one of these, $\Psi_{LH}({\bf x},s)$, is a normalizable zero mode, and
$f(s)=|s|$. We have written the fermions in a two-component notation in
anticipation that these components will correspond to the {\it chiral}
components of $2n$-dimensional spinors. Note that the chirality of the
zero modes is correlated with the sign of the mass term ${\bf M}(s)$

There are also zero modes of the opposite chirality, with momenta near
$\pi/a$, but turning on the Wilson term, $r\neq0$,
 gives them large masses of order
$M_w \approx a r \, ({\pi \over a})^2 \sim {r \pi^2 /a}$.
Thus, in Kaplan's proposal, one arranges that the
doubler modes get large Wilson masses \golterman, while leaving
a chiral zero mode at the wall and a zero mode of the opposite
chirality at the anti-wall.
Of course, the finite volume theory is vector-like, as it must be. But
we expect that in the continuum and infinite volume ($L\to\infty$)
limits, the modes
that are concentrated at the two walls will decouple from each other,
leaving an {\it
effective} $2n$-dimensional {\it chiral} theory on each wall.

We would like to see if this scenario is still realized
when one turns on dynamical gauge fields.
Clearly, this is not going to be a trivial matter. We must ensure that
\item{1)}$2n$-dimensional perturbation theory is successfully reproduced
in the $2n+1$-dimensional theory.
\item{2)}The physics on the two walls is uncorrelated, {\it i.e.}~the
decoupling of the $\psi$'s and the $\chi$'s must persist even in the
presence of dynamical gauge fields.
\item{3)}The nonperturbative physics of the $2n$-dimensional chiral
gauge theory is reproduced successfully, such as fermion number
nonconservation.

Suppose we have $n_L$ fermions with one sign of ${\bf M}(s)$ and $n_R$
fermions with the other sign of ${\bf M}(s)$ in \faction. The free
action \faction\ then has a $U(n_L)\times U(n_R)$ symmetry, and we can
gauge a subgroups $G\in U(n_L)\times U(n_R)$ by replacing
\eqn\gf{
({\bf K} \Psi)_x \to (\hat {\bf K} \Psi)_x \equiv{1 \over a} \sum_{\hat
\mu}^{2n +1}\half \gamma^\mu (U_{\hat \mu} \Psi_{x+ \hat \mu} - U_{-\hat
\mu} \Psi_{x -\hat \mu})+ra (U_{\hat \mu} \Psi_{x + \hat \mu} - 2 \Psi_x +
U_{-\hat \mu} \Psi_{x - \hat \mu})
}
where we have introduced the gauge field through the link variables
$U_\mu ({\bf x}, s) \equiv e^{i a {\bf A}_\mu ({\bf x}, s)} \in G$.

Since the fermion action explicitly breaks the $(2n+1)$-dimensional
(discrete) Euclidean symmetry down to a $2n$-dimensional symmetry, there
is no reason to assume that the gauge action possesses the full
$(2n+1)$-dimensional symmetry. Rather, we introduce an
anisotropic generalization of the Wilson action:
\eqn\sgauge{
S_{\rm gauge} = \sum_{x \in {\bf R}_d} [ \, \beta_{||}(a) \,\,
{\rm Tr} \!\! \prod_{{\bf P}_{x \, ||}} \!\!  U_\mu
+ 2 \beta_\perp(a) \,\, {\rm Tr} \!\!\prod_{{\bf P}_{x \, \perp}}
\!\! U_\mu
+  h.c.].
}
with {\it independent} couplings for plaquettes which lie in $2n$-dimensional
sub-lattices ($\beta_{||})$ and for plaquettes  lying across
$(2n+1)$-th dimension ($\beta_\perp$).

\newsec{Two-Dimensional Axial Schwinger Model}
As a concrete example, in the remainder of this paper,
we consider a two-dimensional
Pythagorean Schwinger model as the simplest chiral
gauge theory.
In Euclidean spacetime, the light cone coordinates are
rotated into a complex coordinate $z \equiv \tau + i x$ and
its complex conjugate.
The model consists of three chiral fermions: two left-handed
$\psi_r, \, \psi_s$ and one right-handed $\psi_t$, and an abelian
gauge field $A_z, \bar A_{\bar z}$ with a field strength
$ \CE \equiv -2i(\partial_z A_{\bar z} - \partial_{\bar z}A_z)$.
Subscripts to the chiral fermions denote electric charge
assignments ${\bf Q} = e (r, s, t)$. We will take the charges to be
commensurate, which while not essential to our arguments, is the
situation which most resembles nonabelian gauge theory in 4 dimensions.
We may then, without loss of
generality, take the charges $r,s,t$ to be integer-valued and ordered as $
0<r<s <t$.
The action is
\eqn\schwaction{
L = {1 \over 2 e^2} {\cal E}^2 + \bar\psi_r(\partial_z
+ r A_z) \psi_r + \bar\psi_s(\partial_z + s A_z) \psi_s
+ \bar\psi_t(\bar \partial_{\bar z} +  t \bar A_{\bar z}) \bar\psi_t\quad.
}
The theory has a $U(1)^3$ symmetry corresponding to independent phase
rotations on the fermions. Denote the corresponding currents by
\eqn\curdef{
\vcenter{\openup1\jot\halign{$#$\hfil\qquad&$#$\hfil\qquad&$#$\hfil\qquad\cr
J^r_z=\bar\psi_r\psi_r&J^s_z=\bar\psi_s\psi_s&J^t_z=0\hfil\cr
J^r_{\bar z}=0\hfil&J^s_{\bar z}=0\hfil&J^t_{\bar z}=\bar\psi_t\psi_t\cr}}.}
One linear combination of these currents is simply the gauge current
$$J_z=r J^r_z +sJ^s_z, \qquad J_{\bar z}=t J^t_{\bar z}.$$
The gauge current is nonanomalous, and the theory consistent, if and
only if the charges satisfy the Pythagorean relation
\eqn\pythagoras{r^2+s^2=t^2\quad.}
The other two linear combinations correspond to {\it global} $U(1)$
symmetries and can be taken to be
\eqn\globcurdef{
\vcenter{\openup1\jot\halign{$#$\hfil\qquad&$#$\hfil\quad\cr
J^L_z=sJ^r_z-rJ^s_z&J^F_z=J^r_z+J^s_z\cr
J^L_{\bar z}=0&J^F_{\bar z}=J^t_{\bar z}\cr}}.}
The first is a symmetry of the quantum theory, for the current is
nonanomalous
$$\partial_{\bar z} J^L_z=0$$
The latter current, which is simply fermion number in this
two-dimensional theory, {\it is} anomalous
\eqn\fviol{
2(\partial_{\bar z}J^F_z+\partial_zJ^F_{\bar z})={1\over \pi} (r+s-t)\CE
}

Anomalous nonconservation of the fermion number is
a notable feature of the Pythagorean Schwinger model.
It is the precise analog of the nonconservation of $B+L$ fermion number
in the $SU(2)_L\times U(1)_Y$ standard model, as discovered by `t~Hooft.
The fermion number anomaly \fviol\ is the two-dimensional
analog of
the $B+L$-number nonconservation
by the weak
instantons \ref\thooft{G. `t Hooft, Phys. Rev. Lett. \bf 37 \rm (1976) 8;
R. Jackiw and C. Rebbi, Phys. Rev. Lett. \bf 37 \rm (1976) 172;
C.G. Callan, R. Dashen and D. Gross,
Phys. Lett. \bf 63B \rm (1976) 334.}.
Fermion number nonconservation has proven to be a sensitive test of
whether a lattice theory correctly reproduces the nonperturbative
physics of chiral gauge theories like the standard model
\refs{\eichtenpreskill,\banks}\ and, as we shall see, the same is true here.

\newsec {Three Dimensional Lattice Gauge Theory:
Continuum \& Infinite Volume Limits}

Having discussed the two-dimensional chiral gauge theory we hope to
reproduce, let us return to the corresponding 3-dimensional lattice theory.
We are interested in exploring the continuum limit
of
\eqn\thraction{
\eqalign{S&=S_{\rm fermion}+S_{\rm gauge}\cr
S_{\rm fermion}&= \sum_{x \in {\bf R}^d } \bar\Psi^r_x({\bf \hat K}+{\bf
M}(s))\Psi^r_x+\bar\Psi^s_x({\bf \hat K}+{\bf
M}(s))\Psi^s_x+\bar\Psi^t_x({\bf \hat K}-{\bf
M}(s))\Psi^t_x\cr
}}
where $S_{\rm gauge}$ is given by \sgauge. This may seem somewhat
formidable, but, at least in perturbation theory, there is an obvious
simplification which we would like to exploit. One can view the
3-dimensional  fermi field $\Psi$ as an infinite collection (on a finite
lattice, actually a finite collection) of 2-dimensional fermi fields.
Usually, this is not a very profitable point of view, but here we note
that {\it almost all} of these fermions have masses of order $M_d$, so
if
we are interested in physics well below this scale, we can integrate
them out, retaining only the zero modes \ezeromodes. In order to do this
in a clean way in the interacting theory \thraction, it behooves us to
partially gauge-fix \thraction, choosing $A_3=0$ gauge\foot{This leaves
a residual gauge-invariance -- the ability to make $x_3$-independent
gauge transformations -- which we will exploit in the following.}. In this
case,
the separation of the light from the heavy fermion modes can be made
cleanly, since the equation determining the zero modes is independent of
the gauge field.  The light fermions are again given by \ezeromodes. Of
course, by working in a particular gauge, our formul\ae\ will not be
{\it manifestly} invariant under the full 3-dimensional gauge symmetry, but
we {\it  will} demand manifest invariance under the group of residual gauge
transformations -- the group of 2-dimensional gauge transformations.
So we will really be considering an {\it effective}
 theory of 2-dimensional fermions
coupled to a 3-dimensional gauge field in $A_3=0$ gauge.

Actually, there is one more term in the gauge action which plays a
crucial role  in realizing the anomalies of the theory \kaplan, namely the
3-dimensional Chern-Simons term, which in $A_3=0$ gauge is simply
\eqn\cs{
{c \over \pi} \int \, d^3 x \, \epsilon^{ij} A_i \partial_3 A_j
}
This term has a {\it calculable} coefficient in the low-energy effective
theory \kaplan: $c=r^2+s^2-t^2$. It vanishes precisely when the
2-dimensional theory is anomaly-free \pythagoras. When the theory {\it
is} anomalous, the nonzero Chern-Simons current
$$\partial_3 J^{CS}_3=-{c\over\pi}\CE$$
compensates for the anomaly in the 2-dimensional gauge current, so that
the full 3-dimensional current is conserved. Note that it is the
Chern-Simons current, not the $x_3$-component of the fermion current
that cancels the anomaly. The fermions are massive off the wall, and
massive degrees of freedom cannot carry off the charge (remember, the
anomaly is a low-energy effect). Rather, it is the gauge-field, which
can propagate off the wall, that, through the Chern-Simons term, carries
off the charge. In a sense, the effective theory is not really
2-dimensional at all; massless degrees of freedom can propagate in the
$x_3$-direction, carrying off the charge.
We can only hope to recover a truly 2-dimensional theory in the
anomaly-free theory, in which the
Pythagorean relation \pythagoras\ is satisfied and the coefficient of
the  Chern-Simons vanishes.

We will, henceforth, assume that \pythagoras\
is satisfied. Also, since we are doing perturbation theory, we will
adopt a continuum language to discuss our calculations. This
is, of course, completely appropriate, since we are interested in
probing the theory at length scales $\gg 1/M_d \gg a$.
Finally, we will first consider the case of a single wall ($L=\infty$),
returning to discuss the finite $L$ corrections later.

Consider sticking a left-handed fermion zero mode
$$\Psi({\bf x},s)=\sqrt{M_d}{\psi({\bf
x})\choose0}\e{-M_d|s|}
$$
into the fermion action. We get
\eqn\sfcont{
\int d s \int d^2{\bf x} \ M_d\ex{-2M_d|s|} \, \bar\psi({\bf
x})\bigl(\partial_z+A_z({\bf x},s)\bigr)\psi({\bf x})
}
Clearly the first term just gives the standard 2-dimensional free
fermion action
$$\int d^2{\bf x}\bar\psi({\bf
x})\partial_z\psi({\bf x})
$$
and leads to the standard 2-dimensional fermion propagator. The second
term represents the coupling of the 2-dimensional fermion to the
3-dimensional gauge field. In momentum space, it is
\eqn\sint{
\int {d^2{\bf q}d^2{\bf k}\over (2\pi)^4}
\int {dk_3\over 2\pi}\, \bar\psi({\bf q})A_z({\bf
k},k_3)\psi(-{\bf q}-{\bf k}) F(k_3)
}
where
\eqn\ff{F(k_3)=\Bigl(1+(k_3/2M_d)^2\Bigr)^{-1}
.}
This interaction is, of course, invariant under the residual gauge
symmetry group of $x_3$-independent (\ie\ 2-dimensional)
gauge transformations.

The action for the gauge field is simply
\eqn\contgauge{\int ds\int d^2{\bf x} {1\over 4e_{||}^2}F_{ij}^2 +
{1\over 2e_\perp^2}(\partial_sA_i)^2
}
which leads to the propagator (in covariant gauge\foot{Recall that we
need to fix the residual 2-dimensional gauge invariance.})
\eqn\prop{D_{ij}({\bf k},k_3)=\left({1\over e_{||}^2}{\bf k}^2+ {1\over
e_\perp^2}k_3^2\right)^{-1}\left[g_{ij}-(1-\xi){k_ik_j\over{{\bf
k}^2+\xi{e_{||}^2\over e_\perp^2}k_3^2}}\right]
}

In a generic Feynman diagram for this theory, we encounter for every
internal photon line a momentum integral of the form
$$\dots\int {d^2{\bf k}\over (2\pi)^2} \int {dk_3\over 2\pi} F(k_3)^2
\gamma^i D_{ij}({\bf k},k_3)\gamma^j \dots
$$
where $F(k_3)$ is the form-factor \ff\ in the interaction vertex \sint.
If we had been doing the same Feynman diagram in the {\it 2-dimensional}
theory, we would have encountered the momentum integral
$$\dots\int {d^2{\bf k}\over (2\pi)^2}\gamma^i {e^2\over{\bf
k}^2}\left[g_{ij}-(1-\xi){k_ik_j\over{\bf k}^2}\right]\gamma^j\dots$$
We can obtain agreement between these two expressions (up to corrections
of order ${\bf k}^2/M_d^2$ or $e^2/M^2_d$) provided we take
\eqn\hier{e_{||}^2\ll M_d\ll e_\perp^2\ll 1/a
}
and identify the 2-dimensional gauge coupling to be
\eqn\coident{e^2=e_{||}^2 M_d/2.
}
With this hierarchy of mass scales we obtain complete agreement between
the perturbation theory of this exotic effective field theory (probed at
energies $\ll M$) and the
genuinely 2-dimensional Pythagorean Schwinger model.

The 3-dimensional theory \thraction\ contained
several dimensionful parameters: the physical size of extra dimension
$L$, the three-dimensional gauge couplings $e_\perp, e_{||}$, the domain
wall mass scale $M_d$, and the lattice spacing $a$. We  might have
enquired at the outset, what combination of these dimensionful
quantities is supposed to be identified as the 2-dimensional gauge
coupling $e^2$? Also, since the 2-dimensional theory has no
dimensionless parameters (which could be formed by ratios of the above
dimensionful ones), what is the hierarchy among these dimensionful
quantities which is necessary to reproduce the 2-dimensional theory?
Equations \hier\ and \coident\ provide the answers to these questions.

At this point, the attentive reader might object to our use of
\contgauge\ as our gauge action. After all, what we have done in
obtaining  \sint\ and \contgauge, is {\it truncate} the original
3-dimensional theory by dropping the heavy fermion degrees of freedom.
This is not the same as integrating them out. Integrating them out
introduces corrections to the effective action for the gauge field \contgauge.
We should show that these corrections do not (qualitatively) change our
results.

The corrections result from doing the 1-loop diagram with  3-dimensional
fermions (with the position-dependent mass term) running around the loop.
This is slightly formidable, but some of the qualitative features can be
seen from doing the same calculation for fermions with a constant mass.
Each term in the effective action comes with a power of $1/M_d$ given by
dimensional analysis. Terms in the effective action with more powers of
the photon field $A_i$, therefore end up having their effects suppressed
by powers of ${\bf k}^2/M_d^2$. The only terms we actually have to worry
about for our low-energy calculation involve two external photon lines,
which correct the photon propagator. We have arranged our fermion
charges so that the parity-violating part of the vacuum polarization
graph cancels, and so the corrections to the photon propagator are all
expressed in terms of a single scalar function $\pi(({\bf
k}^2+k_3^2)/M_d^2)$. The point is that, for the low-energy processes we
are considering, we always have ${\bf k}^2\ll M_d^2$. (Note that it is
crucial here that we are working with a super-renormalizable theory, so
that, even in internal lines, the photon momenta are ``soft".) Thus we
can always neglect the ${\bf k}^2$-dependence of the vacuum polarization
and consider the function $\pi(k_3^2/M_d^2)$. We, of course, {\it
cannot} neglect the $k_3$-dependence, because the typical photon $k_3$
momenta may be of order $M_d$. However, again, up to effects of order ${\bf
k}^2/M_d^2$, this correction to the photon propagator shifts the
constant of proportionality between $e^2$ and $e_{||}^2M_d$ in \coident\
but does not {\it qualitatively} change our conclusions.

What changes when we consider finite $L$, and zero modes
$\psi$ concentrated on the wall and $\chi$ concentrated on
 the anti-wall? Clearly now, a photon
line can couple the fermions $\psi$ on one wall with the fermions $\chi$
on the other. The strength of the coupling is given by the
overlap of the two zero mode wave functions \ezeromodes
$$\int_{-L}^{L} ds\ C\ex{-M_df(s)}\ \!\!\! \cdot C\ex{-M_df(s+L)}
$$
(Previously, both of the wave functions were $\sqrt{M_d}\ex{-M_d|s|}$,
and the integral extended from $-\infty$ to $\infty$.)
Otherwise, the computation is as before. The coupling {\it between}
$\psi$'s and $\chi$'s is exponentially small for large $L$; it goes as
$\ex{-M_dL}$. Thus if we take care to make $L\gg 1/M_d$, the fermions on
the two walls decouple, and, at least in perturbation theory, we have
succeeded in reproducing the 2-dimensional {\it chiral} gauge theory.

\newsec {Nonperturbative Fermion Number Nonconservation}

Finally, we need to return to discuss the fate of the fermion-number
current in the effective theory. We expect it {\it not} to be conserved.
The anomaly equation \fviol\ means that in the theory with dynamical
gauge fields, there are real physical processes which violate fermion
number.

But  3-dimensional lattice theory \thraction\ has fermion number as an
{\it exact symmetry of the lattice action}. Thus the 3-dimensional
fermion number current must be {\it exactly conserved}. Have we arrived
at a paradox?

At first, it appears that the situation resembles the case when the
2-dimensional gauge current is anomalous ($r^2+s^2\neq t^2$). Charge is
not conserved in the would-be 2-dimensional theory. But the
effective theory has a nonzero Chern-Simons term, and the Chern-Simons
current built out of the gauge-field carries off the charge, which is
exactly conserved in the 3-dimensional theory. We conclude that the
theory which wanted to be an anomalous 2-dimensional gauge theory really
isn't 2-dimensional at all. There are light degrees of freedom (built
out of the gauge field) which can carry the charge off into the third
dimension.

But in the nonanomalous theory ($r^2+s^2=t^2$), the coefficient of the
Chern-Simons term vanishes and the theory {\it really is } supposed to
be 2-dimensional. There are no light degrees of freedom to carry the
fermion number off the wall. What is going on?

The answer is, simply, a failure of decoupling. The situation is very
similar to that uncovered by Banks \& Dabholkar \dabholkar. The ``heavy"
fermions that we integrated out to obtain the effective theory of
2-dimensional chiral fermions have zero modes in an instanton
background. Though the fields themselves are massive,
it is a mistake to integrate them out because these zero modes fail to
decouple in an
instanton background. If we insist on integrating them out, they come
back to haunt us by inducing nonlocal terms in the resulting effective
action. These terms restore the $U(1)$ symmetry to an exact symmetry of
the 2-dimensional theory.

A more enlightened (and local!) point of view
is to say that because of this failure of decoupling, the theory isn't
really 2-dimensional at all on the nonperturbative level (\ie\ when we
allow large fluctuations of the gauge field). Because of their zero
modes, it is simply incorrect to
integrate out the ``heavy" fermions, and the theory really
is 3-dimensional.

The theory turns out to be
3-dimensional, not just when the gauge symmetry of the would-be
2-dimensional theory is anomalous, but when
{\it any} of the global symmetries present in the 3-dimensional lattice
theory would be anomalous in the 2-dimensional chiral theory.
Whenever that happens, there is a failure of decoupling which prevents
us from integrating out the ``massive" fermion degrees of freedom to obtain
an effective 2-dimensional theory. When the anomalous symmetry is a gauge
symmetry, the failure is seen in perturbation theory -- due
to the presence of a nonvanishing Chern-Simons term. When it is a global
symmetry, the failure manifests itself only nonperturbatively -- when ``large"
fluctuations of the gauge field are taken into account.

The dilemma faced by this theory is a stark one. Fermion number is an
exact symmetry of the lattice action. This means that {\it either} there
are light degrees of freedom that carry the fermion number off of the
wall -- in which case, it is inconsistent to think of the effective
theory as 2-dimensional-- {\it or} the fermion number symmetry is also
an exact symmetry of the 2-dimensional theory. Either way, we do not
obtain the physics that we want: a 2-dimensional theory with fermion
number violation.

The lesson that we should draw from this analysis is that for proposal
for putting chiral fermions on the lattice must not only correctly
reproduce the gauge anomalies of the would-be chiral gauge theory, it
must also reproduce the anomalies in whatever {\it global} symmetries
may be present in the model (like $B+L$ in the standard model). This is
a tall order, and so far, neither Kaplan's model nor any other proposal
\refs{\wilsonyukawa, \eichtenpreskill, \staggered}\
seems to be up to the task.

\leftline {\bf Acknowledgment}\hfill\break
We are grateful to D.~Kaplan whose seminar stimulated our
interest to this problem, and to T.~Banks, C.G.~Callan,
J.~March-Russell,  J.~Preskill and F.~Wilczek
for useful discussions. We would also like to gratefully acknowledge the
hospitality of the ITP at UC Santa Barbara, where this manuscript was
written.

\listrefs
\bye
\end